\documentclass[11pt,letterpaper]{article}
\usepackage{amssymb,amsmath,graphicx}
\usepackage[round]{natbib}

\newcommand{\D}{\ensuremath{\mathcal{D}}}
\newcommand{\fud}[1][]{\frac{1}{2 #1}}
\DeclareMathOperator{\Ei}{Ei}%
\DeclareMathOperator{\Pf}{Pf}%
\newcommand{\expp}[1]{ \mathop\mathit{e}\nolimits^{#1}}
\newcommand{\dert}[3][]{\frac{\mathrm{d}^{#1} #2}{\mathrm{d} #3^{#1}}}
\newcommand{\derp}[3][]{\frac{\partial^{#1} #2}{\partial #3^{#1}}}
\newcommand{\derpp}[3]{\frac{\partial^2 #1}{\partial #2 \partial #3}}
\newcommand{\ud}[2][]{\textrm{d}^{#1}{#2}\,}
\newcommand{\uD}[1]{\D{#1}\,}
\newcommand{\vd}[2][]{\,\textrm{d}^{#1}{#2}}

\newcommand{\Eqref}[1]{Eq.~\eqref{#1}}

\newcommand{\ie}{\emph{i.e.}}
\newcommand{\eg}{\emph{e.g.}}
\newcommand{\av}[1]{\left\langle #1 \right\rangle}
\newcommand{\Es}{{\varepsilon_\mathrm{s}}}
\newcommand{\Eo}{{\varepsilon_0}}
\newcommand{\PB}[2]{\{#1,#2\}_\mathrm{\scriptscriptstyle PB}}
\newcommand{\fini}{f_\text{i}}

\allowdisplaybreaks[3]

\begin{document}

\title{\textbf{\Large Activation-like processes at zero temperature}}

\linespread{0.93}
\author{
    \normalsize Daniel Arteaga\thanks{Electronic address: \texttt{darteaga@ffn.ub.es}.}$\ ^{,\mathrm a}$,
    Esteban Calzetta\thanks{Electronic address: \texttt{calzetta@df.uba.ar}.}$\ ^{,\mathrm b}$,
    Albert Roura\thanks{Electronic address: \texttt{roura@physics.umd.edu}.}$\ ^{,\mathrm c}$, \\
    \normalsize and Enric Verdaguer\thanks{Electronic address: \texttt{verdague@ffn.ub.es}.}$\ ^{,\mathrm a}$\\[1.5mm]
    \normalsize \itshape $^\mathrm a$Departament de F\'\i sica Fonamental \\
            \normalsize \itshape and CER en  Astrof\'\i sica, F\'\i sica de Part\'\i cules i Cosmologia, \\
            \normalsize \itshape Universitat de Barcelona, Av. Diagonal 647, 08028 Barcelona, Spain. \\[1.5mm]
        \normalsize \itshape $^\mathrm b$Departamento de F\'\i sica, Universidad de Buenos Aires, \\
            \normalsize \itshape Ciudad Universitaria, 1428 Buenos Aires, Argentina. \\[1.5mm]
        \normalsize \itshape $^\mathrm c$Department of Physics, University of Maryland, \\
             \normalsize \itshape College Park, Maryland 20742--4111.
}
\date{}

\maketitle \linespread{1}

\vspace*{-1.5\baselineskip}

\begin{abstract}
We examine the possibility that a metastable quantum state could
experiment a phenomenon similar to thermal activation but at zero
temperature. In order to do that we study the real-time dynamics
of the reduced Wigner function in a simple open quantum system: an
anharmonic oscillator with a cubic potential linearly interacting
with an environment of harmonic oscillators. Our results suggest
that this activation-like phenomenon exists indeed as a
consequence of the fluctuations induced by the environment and
that its associated decay rate is comparable to the tunneling rate
as computed by the instanton method, at least for the particular
potential of the system and the distribution of frequencies for
the environment considered in this paper. However, we are not able
to properly deal with the term which leads to tunneling in closed
quantum systems, and a definite conclusion cannot be reached until
tunneling and activation-like effects are considered
simultaneously.
\end{abstract}

\section{Introduction}

The study of the decay rate of a state trapped in a metastable
state by a potential barrier has a long and distinguished history
both in statistical physics and in quantum mechanics. In
statistical mechanics one is usually worried about the thermal
activation effect, by which a particle escapes over a potential
barrier due to the fluctuations induced by a thermal bath. As a
paradigm we have the classical work by \cite{Kramers40}, who
considered a classical Brownian particle trapped in a metastable
minimum and computed the escape probability by analyzing the
dynamics of its probability distribution function, both in the
underdamped and overdamped cases. In quantum mechanics one is
interested in the tunneling effect, by which a particle escapes
the local minimum by traversing the potential barrier through a
classically forbidden region. A technique which has been of great
success to compute the tunneling rate in quantum mechanics and
quantum field theory is the instanton method
\citep{Coleman77,CallanColeman77,Coleman}, where the decay
probability can be computed in terms of the classical trajectories
in imaginary time. In both thermal activation and tunneling the
decay rate $r$, meaning the decay probability per unit time,
follows an approximate exponential law, $r \approx A \exp{(-B)}$.
In thermal activation, $B=V_\mathrm s/(k T)$, where $k$ is
Boltzmann's constant, $T$ is the absolute temperature and
$V_\mathrm s$ is the height of the free energy measured from the
metastable minimum; in tunneling $B=S_\mathrm E/\hbar$, where
$S_\mathrm E$ is the action for a suitable trajectory which goes
under the barrier in imaginary time.

In recent years mesoscopic physics has become a center of
attention. Experimental advances are pushing the boundaries
between classical and quantum systems and lead, in particular, to
the possibility of observing quantum tunneling for systems that
can be described by macroscopic variables. These are essentially
open quantum systems, which are characterized by a distinguished
subsystem within a larger closed quantum system, described by some
degrees of freedom which are subject to physical experimentation,
and the rest of the system, described by generally unobservable
degrees of freedom which act as an environment or bath for the
distinguished subsystem. The environment induces both dissipation
and noise to the distinguished system, which is usually referred
to as the ``system'' for short. Many of these open quantum systems
can be equivalently represented by a particle subject to an
arbitrary external potential and coupled to an environment
consisting of an infinite set of independent harmonic oscillators.
A number of known physical systems can be modelled by adjusting
the coupling of the system and environment variables and choosing
appropriate potentials.

\cite{CaldeiraLeggett81,CaldeiraLeggett83a}, in two influential
papers, considered the effect of an environment on quantum
tunneling. They were able to generalize the instanton method to a
simple open quantum system. In particular, they considered an
anharmonic quadratic plus cubic potential bilinearly coupled an
environment at zero temperature consisting of an infinite set of
harmonic oscillators, with frequencies distributed according to
the so-called ohmic distribution. They argued that this system is
a very good model for the flux trapped in a superconducting
quantum interference device (SQUID), a single Josephson junction
biased by a fixed external current, and others. Assuming that the
environment degrees of freedom are only weakly perturbed by the
interaction with the system, they concluded that dissipation
always tends to suppress tunneling.

\cite{FujikawaEtAl92a,FujikawaEtAl92b} reanalyzed the same problem
as Caldeira and Leggett employing canonical perturbation theory
and the second quantization formalism for the system instead of
the instanton method, with a quadratic plus quartic potential in
order to have a well-defined ground state. By considering the two
lowest energy eigenstates of the system they were able to
reproduce Caldeira and Leggett's results. They studied the effect
of the next two excited states and concluded that, once these
states are taken into account, dissipation can enhance tunneling
for some distributions of frequencies of the environment, but not
for the ohmic distribution considered by Caldeira and Leggett.

In some recent papers a different real-time approach to compute
the vacuum decay rate in quantum field theory was introduced
\citep{CalzettaRouraVerdaguer01,CalzettaRouraVerdaguer02} as a
first step to consider situations far from the equilibrium. The
analysis is based on the time evolution equation, the so-called
master equation, for the reduced Wigner function describing the
open quantum system. The quantum field theory problem was reduced
to an open quantum system described by a single degree of freedom
associate to the modes of the field which are nearly homogeneous
within a region whose size corresponds to a nucleating bubble of
true vacuum, and coupled to an infinite set of harmonic
oscillators corresponding to the inhomogeneous modes of the field.
The coupling was linear in the system variables but quadratic in
the environment ones. The master equation contained dissipation
and noise terms, which describe the influence of the environment
on the system, as well as derivative terms  in the momentum
coordinates (third order for a cubic potential), which are
responsible for quantum tunneling in closed quantum systems.

Unfortunately it is not possible to compute the total decay rate
in a closed analytic or semianalytic form because it is very
difficult to deal with the third order derivative terms. One
should resort to methods such as those based on matrix continued
fractions in order to compute the decay rate master equations with
third-order derivative terms \citep{RiskenVogel88}. However in the
work by \cite{CalzettaRouraVerdaguer01,CalzettaRouraVerdaguer02},
as well as in the present paper, one was primarily interested in
the contribution from the environment backreaction on the vacuum
decay rate and, hence, the rather drastic approximation of
neglecting the third derivative term responsible for tunneling was
made. The decay rate obtained was entirely due to the terms which
at high temperature are responsible for thermal activation. In
principle this approximation should be correct provided that the
timescales associated to the activation-like effect and the
tunneling effect are very different. The remarkable result was
that the activation-like effect produced by the backreaction of
the inhomogeneous modes was, in fact, larger that the tunneling
effect obtained with the instanton method.

The fact that activation may be important as a backreaction effect
can be understood by noting that the characteristic timescale for
the decay process is much larger than the dynamical and relaxation
timescales; see \cite{CalzettaVerdaguer99} for a detailed
analysis. In fact, although the effects of dissipation and noise
are very small on the characteristic dynamical timescale, they can
have a cumulative effect which becomes important in the long run.
The real time approach based on the master equation seems a
suitable technique to deal with those backreaction effects, but is
difficult to implement when addressing the tunneling effect. On
the other hand, the instanton method is very well suited for
studying the tunneling effect, but, at least in some cases, seems
to underplay the backreaction of the environment on the system.

The aim of this paper is to study whether the results obtained by
\cite{CalzettaRouraVerdaguer01,CalzettaRouraVerdaguer02} are a
particular feature of the particular system that they considered
or, on the contrary, a general feature which can be extended to a
wider class of systems. Therefore, using the real-time techniques
mentioned above, we reanalyze the simpler model studied by
\cite{CaldeiraLeggett81,CaldeiraLeggett83a}, in order to check
whether a similar activation-like effect is also found in that
case.

There are several differences between the model considered in this
paper and the field theory case analyzed by
\cite{CalzettaRouraVerdaguer01,CalzettaRouraVerdaguer02}. First,
we  consider here a bilinear coupling between the system and
environment, whereas the coupling considered in the field theory
case was quadratic in the environment degrees of freedom. Second,
in the field theory case both the spectral distribution of the
environment frequencies and the value of the coupling constant
were \emph{a priori} determined by the particular
system-environment separation considered there, whereas we shall
freely choose the spectral distribution and the coupling
parameter. Third, we will employ several techniques (harmonic
approximation, Kramers method and lowest eigenvalue expansion)
which could not be applied to the field theory problem because of
the particular features of that model.

The plan of the paper is the following. In section 2 we set up the
model. In section 3, we present the master equation for the
reduced Wigner function. Then, following closely
\cite{CalzettaRouraVerdaguer02}, we neglect the third order
derivative term and concentrate on the weak dissipation case by
studying the averaged dynamics over an oscillation period. In
section 4 we obtain an analytical expression for the decay rate by
assuming a harmonic approximation for the classical trajectories.
Two alternative approaches are employed. The first one is based on
a perturbative expansion for the lowest eigenvalue, whereas the
second one is based on Kramers's classical work. Finally, in
section 5 we discuss the implications of our result.

\section{The open quantum system model}

Let us consider a particle of mass $M$, the ``system'', subject to
an arbitrary potential $V(x)$ and coupled to a bath of independent
harmonic oscillators of mass $m$, the ``environment''. Let us
assume that the system and the environment are linearly coupled.
The action for the whole set of degrees of freedom is given by
\begin{subequations}
\begin{equation}
    S[x,\{q_j\}] = S_\text{s}[x] + S_\text{e}[\{q_j\}] + S_\text{int}[x,\{q_j\}],
\end{equation}
where the terms on the right-hand side, which correspond to the
action of the system, the environment and the interaction term
respectively, are given by
\begin{align}
    S_\text{s}[x] &= \int \ud t \left( \fud M \dot x^2 - V(x)\right), \\
    S_\text{e}[\{q_j\}] &= \sum_j \int \ud t \left( \fud m \dot q_j^2- \fud
    m \omega_j^2q_j^2\right), \\
\begin{split}
    S_\text{int}[x,\{q_j\}] &= \sum_j c_j \int \ud{t} x(t) q_j(t),
\end{split}
\end{align}
\end{subequations}
with $c_j$ being the system-environment coupling parameters and
$\omega_j$ the environment oscillator frequencies. The system
potential $V(x)$ includes a quadratic part, corresponding to an
oscillator of frequency $\Omega_0$, and an anharmonic part
$V^\text{(nl)}$,
\begin{equation}
    V(x) = \fud \Omega_0^2 x^2 + V^\text{(nl)}(x).
\end{equation}
At this point, the potential $V^\text{(nl)}(x)$ is arbitrary, but
later on we will take a cubic potential $V^\text{(nl)}(x)= -
(\lambda/6)x^3$. It will be convenient for us to rewrite the
interaction term as
\begin{equation}
    S_\text{int}[x,\{q_j\}] = \int_0^\infty \frac{2m\omega}{\pi c(\omega)} I(\omega) \int
    \ud t x(t) q(t;\omega),
\end{equation}
where  $c(\omega)$ and $q(t;\omega)$ are functions such that
$c(\omega_j)=c_j$ and $q(t;\omega_j)=q_j(t)$ and
\begin{equation}
    I(\omega)= \sum_j \frac{\pi c_j^2}{2m\omega_j}
    \delta(\omega-\omega_j)
\end{equation}
is the spectral density of the environment.

When the system and the environment are initially uncorrelated,
\ie, when the initial density matrix factorizes, the evolution for
the reduced density matrix can be written as
\begin{equation} \label{PropJ}
    \rho_r(x,x',t) = \int \ud{x_\mathrm{i}} \ud{x'_\mathrm{i}}
    J(x,x',t;x_\mathrm{i},x'_\mathrm{i},t_\mathrm{i}) \rho_r(x_\mathrm{i},x_\mathrm{i}',t_\mathrm{i}),
\end{equation}
where the propagator $J$ is found to be, in a path integral
representation,
\newcommand{\SIF}{S_\text{IF}}
\begin{equation}
    J(x_\mathrm{f},x'_\mathrm{f},t;x_\mathrm{i},x'_\mathrm{i},t_\mathrm{i})  =
    \int \limits_{x(t_\mathrm{i})=x_\mathrm{i}}^{x(t)=x_\mathrm{f}} \uD x
    \int \limits_{x'(t_\mathrm{i})=x'_\mathrm{i}}^{x'(t)=x'_\mathrm{f}} \uD x'
    \expp{i\left( S[x] - S[x']+
    \SIF[x,x']\right)/\hbar}
\end{equation}
where $\SIF[x,x']$ is the influence action, related to the
influence functional $F_\text{IF}$ introduced by
\cite{FeynmanVernon63} through $F_\text{IF}[x,x']=\exp{(i
\SIF[x,x']/\hbar)}$. For a Gaussian initial density matrix for the
environment, the influence action can be expressed as
\citep{FeynmanVernon63,FeynmanQMPI,CaldeiraLeggett83b}:
\begin{equation}
\begin{split}
    \SIF[x,x'] = &-2 \int_{t_\mathrm{i}}^{t} \ud s \int_{t_\mathrm{i}}^s \ud {s'}
    \Delta(s) D(s,s') \Sigma(s') \\ &+ \frac{i}{2}
    \int_{t_\mathrm{i}}^{t} \ud s \int_{t_\mathrm{i}}^{t} \ud {s'}
    \Delta(s)N(s,s')\Delta(s'),
\end{split}
\end{equation}
where $\Sigma \equiv (x+x')/2$ and $\Delta \equiv x'-x$. The
kernels $D(t,t')$ and $N(t,t')$ are called the dissipation and
noise kernels, respectively. If initially we are at thermal
equilibrium at a temperature $T$ these kernels are given by:
\begin{subequations}
\begin{align}
    D(t,t') &= \int _0^\infty \frac{\ud \omega}{\pi} I(\omega) \sin
    \omega(t-t'),\\
    N(t,t') &= \int _0^\infty \frac{\ud \omega}{\pi} I(\omega)
    \coth \left(\frac{\hbar \omega}{2kT}\right)
    \cos \omega(t-t').
\end{align}
\end{subequations}

The influence action can be divergent and a renormalization
procedure may be required, as can be seen by reexpressing the
influence action as
\begin{equation}
\begin{split}
    \SIF[x,x'] = & \int_{t_\mathrm{i}}^{t} \ud s \int_{t_\mathrm{i}}^t \ud {s'}
    \Delta(s) H(s,s') \Sigma(s') \\ &+ \frac{i}{2}
    \int_{t_\mathrm{i}}^{t} \ud s \int_{t_\mathrm{i}}^{t} \ud {s'}
    \Delta(s)N(s,s')\Delta(s'),
\end{split}
\end{equation}
where, at least formally, $H(t,t')\equiv -2 \theta(t-t') D(t,t')$,
being $\theta(t-t')$ the step function. The kernel $H(t,t')$ is a
product of two distributions, which in general is not well defined
and may contain divergences. Nevertheless, it is always possible
to introduce suitable counterterms in the bare frequency of the
system $\Omega_0$ in order to compensate the divergent terms
coming from $H(t,t')$. See
\cite{CaldeiraLeggett83b,RouraVerdaguer99,CalzettaRouraVerdaguer00}
for more details. However in the particular problem in which we
are interested this issue will turn to be unimportant since the
divergent parts of the kernel $H(t,t')$ will cancel in the final
results. Thus, we can use the bare kernel $H(t,t')$ (with some
implicit regularization) instead of its renormalized expression.

Following \cite{CaldeiraLeggett81,CaldeiraLeggett83a} we shall
consider the case of zero temperature and ohmic environment, in
which we have a continuum of harmonic oscillators in the
environment distributed according to:
\begin{equation}
    I(\omega) = \eta \omega.
\end{equation}
With this spectral density the expectation value of $\hat x(t)$
obeys the equation of motion of a classical damped oscillator with
a friction coefficient given by the proportionality constant
$\eta$. In this case dissipation and noise kernels are found to
be:
\begin{subequations}
\begin{align}
    D(t,t') &= \eta \delta'(t-t'), \\
    N(t,t') &= \frac{\eta}{\pi} \Pf \frac{-1}{(t-t')^2},
\end{align}
\end{subequations}
where $\Pf$ indicates the Hadamard finite part prescription
\citep{Schwartz}. Later on we will need the expressions of the
Fourier transforms of the noise and dissipation kernels,
\begin{subequations}
\begin{align}
    D(t,t') = \int \frac{\ud{\omega}}{2\pi} \expp{-i\omega(t-t')}
    \tilde D(\omega), \\
    N(t,t') = \int \frac{\ud{\omega}}{2\pi} \expp{-i\omega(t-t')}
    \tilde N(\omega),
\end{align}
\end{subequations}
which in the case of zero temperature and ohmic environment are
given by
\begin{equation}
\tilde D(\omega) = i \eta \omega, \qquad \tilde N(\omega) = \eta
|\omega|.
\end{equation}

\begin{figure}[!ht]
    \centering
    \includegraphics[width=0.60\textwidth]{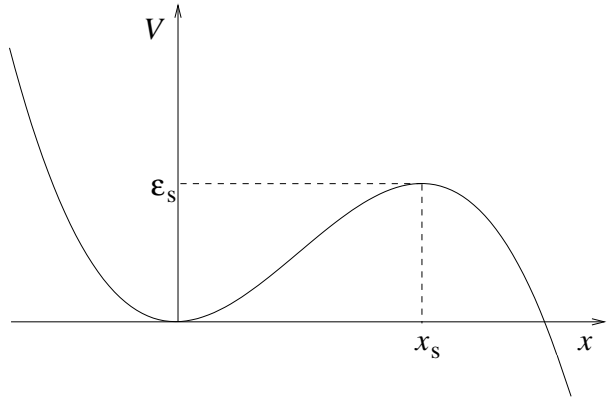}
    \caption{Plot of the potential $V(x)$ under which the particle
    is confined. The maximum of the potential barrier is at $x_\mathrm s$ and corresponds to an
    escape energy $\Es$, which we consider much larger than the zero point energy of the harmonic
    oscillator, $\hbar \Omega_0/2$. }\label{figure}
\end{figure}

We will concentrate on the potential $V(x)=(1/2)M\Omega_0^2 x^2 -
(\lambda/6)x^3$, which exhibits a metastable minimum at $x=0$ and
an unstable maximum at $x=x_\text{s} = 2 M \Omega_0^2/\lambda$,
which corresponds to an energy $\Es=V(x_\text{s})=2 M^3
\Omega_0^6/(3\lambda^2)$ (see Fig. \ref{figure}). The system state
will be peaked located around the metastable minimum at $x=0$, and
can escape through the potential barrier. We will consider that
once the particle exits the potential well region, it never
reenters. Since this potential is not bounded from below, it
should be understood as an approximation to a more realistic
situation in which there exists an absolute minimum, located at a
much lower energy, so that the return probability is negligible.
We will restrict to the situation in which the energy barrier is
much larger than $\hbar \Omega_0$.

\section{Phase-space dynamics}

\subsection{Evolution of the reduced Wigner function}

The reduced Wigner function $W_\mathrm{r}$ is a phase space
distribution defined from the reduced density matrix
$\rho_\mathrm{r}$ by the following integral transform:
\begin{equation}
    W_\text{r}(x,p,t) = \frac{1}{2\pi \hbar} \int_{-\infty}^{\infty}
    \ud{\Delta} \expp{i p \Delta/\hbar}
    \rho_\text{r}(x-\Delta/2,x+\Delta/2,t).
\end{equation}
The Wigner function is a quantum mechanical analogue of a phase
space probability distribution \citep{Wigner32, HilleryEtAl84}.
The partial distribution $\int_{-\infty}^\infty \ud{p}
W_\mathrm{r}(x,p,t)$ gives the probability density of finding the
system at the position $x$; in the same way,
$\int_{-\infty}^\infty \ud{x} W_\mathrm{r}(x,p,t)$ gives the
probability density of finding the system with momentum $p$.
However, the uncertainty principle prevents us from determining at
the same time the position and the momentum of a particle, so that
the Wigner function cannot be interpreted as a true probability
density in phase space. In fact, the Wigner function it is not
necessarily positive defined everywhere and in general it may
acquire negative values.

Up to terms of order $c^2_j$, $\hbar c^2_j$, and $\hbar^2$, with
$c_j$ being the environment coupling constants, the reduced Wigner
function $W_\mathrm{r}=W_\mathrm{r}(x,p,t)$ obeys the following
evolution equation \citep{CalzettaRouraVerdaguer02,RouraThesis},
which we shall call master equation:
\begin{equation} \label{Master}
    \derp{W_\mathrm{r}}{{t}} =  \PB{H_\mathrm{s}}{W_\mathrm{r}} + \derp{}{p}\left(\mathcal D W_\mathrm{r} + \hbar \PB{\mathcal N}{W_\mathrm{r}}\right) - \hbar^2
    \frac{\lambda}{24} \derp[3]{W_\mathrm{r}}{p},
\end{equation}
where $\PB{\cdot}{\cdot}$ are the Poisson brackets,
\[
    \PB{f}{g}=\derp{f}{x}\derp{g}{p} -\derp{f}{p}\derp{g}{x},
\]
$H_\mathrm{s}$ is the reduced system Hamiltonian,
\begin{equation}
    H_\mathrm{s} = \frac{p^2}{2M} + \fud M\Omega_0^2 x^2 -
    \frac{\lambda}{6} x^3,
\end{equation} and $\mathcal D$ and $\mathcal N$ are given by
\begin{subequations}
\begin{align}
    \mathcal D(t) &= -2\int_{t_\mathrm{i}}^{t} \ud{t'} D(t,t')
    X^{}(t';x,p), \label{Drara}\\
    \mathcal N(t) &= \int_{t_\mathrm{i}}^{t} \ud{t'} N(t,t') X(t';x,p),
    \label{Nrara}
\end{align}
\end{subequations}
where $X^{}(t';x,p)$ is a solution of the classical equations of
motion associated to the hamiltonian $H_\mathrm{s}$,
\begin{subequations}\label{ClasMot}
\begin{align}
    \dert{X}{t'} &= \frac{P}{M},\\
    \dert{P^{}}{t'} &= - M \Omega_0^2 X^{}
    + \lambda \frac{X^2}{2}, \label{ClasMotb}
\end{align}
\end{subequations}
with final conditions $X(t)=x$ and $P(t)=p$.

For the case of ohmic environment the friction coefficient $\eta$
is proportional to $c_j^2$, and the above approximation for the
evolution equation will be valid and consistent when the condition
$S \gg \hbar \gtrsim S \gamma/\Omega_0$ is fulfilled, where
$\gamma \equiv \eta/(2M)$ is the characteristic dissipation
frequency and $S$ is the typical action for the process
considered. If we are interested in studying the jump over a
potential barrier of height $\Es$ this condition can be restated
as $\Es \gg \hbar \Omega_0 \gtrsim \Es \gamma/\Omega_0$, which
shows that the master equation will be valid for small dissipation
and large energy barriers.

At this point it is worth making a comment on the notation.
Throughout this paper we will use lowercase letters
($x,p,\theta,j\ldots$) to indicate phase-space variables, which
are the arguments of phase-space distributions such as the Wigner
function, whereas the corresponding uppercase letters
($X,P,\Theta,J\ldots$) will indicate time-functions which give the
phase-space position of a particle in a given time.

In order to introduce the Fourier transform of the coefficients
$\mathcal D$ and $\mathcal N$ later on, it will be convenient for
us to replace the integration limit in Eqs.~\eqref{Drara} and
\eqref{Nrara} by $-\infty$, although the initial conditions are
set up at $t_\mathrm i=0$. For the coefficient $\mathcal D$ this
does not introduce any error, since the dissipation kernel only
has support in $t=t'$. For the coefficient $\mathcal N$, this
introduces a small error, which can be estimated by performing the
integral in \Eqref{Drara}, choosing a periodic function for
$X(t')$ of frequency $\Omega_0$; this will be enough in the
context of the adiabatic approximation, which we will introduce
later on. The result of the calculation shows that for times $t$
which verify $t \gg \Omega_0^{-1}$ the contribution of the
integral from $-\infty$ to $0$ is comparatively small. Since at
the end we shall be interested in studying time scales of the
order of the decay time, which are much larger than the
characteristic dissipation time $\gamma^{-1}$, which will in turn
be much larger than the time of oscillation $\Omega^{-1}_0$, the
approximation can be considered safe.

If the system were isolated, \Eqref{Master} would reduce to
\begin{equation} \label{VonNeumann}
    \derp{W}{{t}} = \PB{H_\mathrm{s}}{W}- \hbar^2 \frac{\lambda}{24}
    \derp[3]{W}{p},
\end{equation}
where $W$ is the Wigner function of the closed system. This
equation is exactly equivalent to von Neumann's equation for the
density matrix of a one dimensional quantum mechanical system with
the potential $V(x)$. If the last term in this equation were not
present, the evolution of the Wigner function would be entirely
equivalent to that of a classical ensemble in the phase space.
Hence, the term with the third derivatives must be the responsible
for tunneling. In principle, one could compute the tunneling
amplitude from \Eqref{VonNeumann}, but in practice tunneling is
more easily calculated with the WKB approximation to the
Schr\"odinger equation \citep[see,
\eg,][]{GalindoPascual,LandauQM} or the instanton method
\citep{Coleman77,CallanColeman77,Coleman}. As we have discussed,
in this contribution we are not going to deal with tunneling, but
rather to compute the effect due to activation. With this aim, we
shall neglect the last term of \Eqref{Master}, which is the
responsible for tunneling. Hence we will use the following
equation for the distribution function $W_\mathrm{r}$:
\begin{equation} \label{FP}
    \derp{W_\mathrm{r}}{{t}} = \PB{H_\mathrm{s}}{W_\mathrm{r}} + \derp{}{{p}}\left(\mathcal D W_\mathrm{r} +
    \hbar \PB{\mathcal N}{W_\mathrm{r}}\right).
\end{equation}

Formally \Eqref{FP} can be thought as a Fokker-Planck equation,
describing the dynamics of an ensemble of points in the phase
space.  The dynamics of this ensemble of points can be equally
characterized by means of the following Langevin equation:
\begin{subequations} \label{Langevin}
\begin{align}
    \dot X_\mathrm{s} &= \frac{P_\mathrm{s}}{M},\\
\begin{split}
    \dot P_\mathrm{s} &= -V'(X_\mathrm{s}) - \int_{-\infty}^\infty \ud t' H(t,t') X_\mathrm{s}(t')
    + \xi \\
    &= - M \Omega_0^2 X_\mathrm{s} - \eta \dot X_\mathrm{s}
    + \lambda \frac{X_\mathrm{s}^2}{2} + \xi ,
\end{split}\label{LangB}
\end{align}
where $\xi$ is a Gaussian noise of zero mean and correlation
function
\begin{equation}
    \av{ \xi(t) \xi(t')}_\xi = \hbar N(t,t'),
\end{equation}
\end{subequations}
$X_\mathrm s=X_\mathrm s(t,\xi]$ and $P_\mathrm s=P_\mathrm s
(t,\xi]$ are the stochastic functions corresponding to the phase
space variables $x$ and $p$ respectively, and the second equality
in \Eqref{LangB} is just valid in the case of ohmic environment.
This Langevin equation does not describe actual trajectories of
the system (meaning a continuous sequence of projectors for the
position and the momentum of the system at each instant of time,
which would violate Heisenberg's uncertainty principle)  but must
be regarded as a formal computational tool, in the same way as the
Wigner function does not correspond to a true probability density.
In fact, Langevin-like equations appear naturally in the context
of open quantum systems when trying to derive the dynamics of the
reduced Wigner function from the path integrals in \Eqref{PropJ}
\citep{CalzettaRouraVerdaguer00}.

\subsection{Action-angle variables}

In order to obtain explicit expressions for the coefficients
$\mathcal D$ and $\mathcal N$, we need to solve the set of
Eqs.~\eqref{ClasMot}, which describe the motion of a classical
particle in the potential $V(x)$. Since we are interested in the
motion inside the potential well, which is periodic, it is
possible to introduce action-angle variables $\theta$ and $j$
\citep[see, \eg,][]{Goldstein}. The action variable is defined by
\begin{equation}
    j = \frac{1}{2\pi} \oint P \vd{X},
\end{equation}
whereas the angle variable $\theta$ changes from zero to $2\pi$
and is canonically conjugate to $j$. Recall that the reduced
system Hamiltonian can be entirely written in terms of the action
variable, $H_\text{s}=H_\text{s}(j)$.

We shall consider $\theta$ and $j$ phase-space variables like $x$
and $p$, and we will analyze the Fokker-Planck equation in terms
of these new variables. However, the solution of
Eqs.~\eqref{ClasMot} can be also described by giving the
trajectory of the particle in the $\theta$--$j$ space. In a
completely analogous way to $X$ and $P$, we will consider the
functions $\Theta(t';\theta,j)$ and $J(t';\theta,j)$, which give
the angular position and action of a particle satisfying the set
of Eqs.~\eqref{ClasMot}, with final conditions $\Theta(t)=\theta$
and $J(t)=j$. The functions $\Theta$ and $J$ satisfy the equations
of motion $\dot \Theta = \omega(J)$ and $\dot J=0$, where
$\omega(j) = \mathrm{d}H_\text{s}(j)/\mathrm{d}j$ is the frequency
of oscillation. With the aforementioned boundary conditions, these
equations of motion can be immediately solved to give
$\Theta(t';\theta,j) = \theta + \Omega(j)(t'-t)$ and
$J(t';\theta,j)=j$.

Since $\theta$ is an angle, the transformation equation
$x=x(\theta,j)$ is periodic in $\theta$,
$x(\theta,j)=x(\theta+2\pi,j)$ and thus it can be decomposed in
terms of a Fourier series with respect to $\theta$,
\begin{equation} \label{xthetaj}
    x(\theta,j) = \sum_{n=-\infty}^{\infty} \expp{in \theta}
    x_n(j),
\end{equation}
where $x_{-n}(j)=x^*_n(j)$ since $x$ is real. The trajectory of
the particle can be also  decomposed in terms of the Fourier
series associated to the angular coordinate:
\begin{equation} \label{Xthetaj}
\begin{split}
    X(t';\theta,j) &= \sum_{n=-\infty}^{\infty} \expp{in \Theta(t';\theta,j)}
    x_n(J(t';\theta,j))\\
     &= \sum_{n=-\infty}^{\infty} \expp{in [\theta + \Omega_0(t'-t)]}
    x_n(j).
\end{split}
\end{equation}
Then we can write the functions $\mathcal D(t)$ and $\mathcal
N(t)$ appearing in \Eqref{FP} as
\begin{subequations}
\begin{align}
    \mathcal D(t) &=  -2\int_{-\infty}^{t} \ud{t'} D(t,t') X(t')
    = \sum_{n=-\infty}^{\infty}  \expp{in \theta} x_n(j) D_n(j), \\
    \mathcal N(t) &= \int_{-\infty}^{t} \ud{t'} N(t,t') X(t')
    = \sum_{n=-\infty}^{\infty}  \expp{in\theta} x_n(j) N_n(j),
\end{align}
\end{subequations}
where $D_n(j)$ and $N_n(j)$ are given by
\begin{subequations}
\begin{align} \label{DnNn}
    D_n(j)&= \int \frac{\ud\omega}{2\pi}
    \frac{2i \tilde D(\omega)}{\omega + n\Omega(j) - i \epsilon},
    \\
    N_n(j)&= \int \frac{\ud\omega}{2\pi}
    \frac{- i \tilde N(\omega)}{\omega + n\Omega(j) - i \epsilon}
\end{align}
\end{subequations}
In order to derive these last expressions we have made use of the
following equality: $\int_0^\infty \exp (isu) = i/(s+i\epsilon)$,
where $\epsilon$ is an arbitrarily small positive real number.

\subsection{Weak dissipation limit: averaging over angles}

As discussed by \cite{Kramers40}, in the case of small
dissipation, \ie, $\gamma \ll \Omega_0$, the phase space dynamics
will mostly correspond to a gradual change of the distribution of
the ensemble over the different energy values. The change of the
Wigner function over an oscillation period will be small, so that
we may suppose that the reduced Wigner function only depends on
the action variable $j$ (or the energy $E$), and does not depend
on the angular variable $\theta$, $W_\text{r}(\theta,j)=F(j)$.
Thus, we can obtain a simpler equation by averaging all the terms
of the Fokker-Planck equation over the variable $\theta$.
Furthermore, in this case the averaged Wigner function $F$ is a
partial distribution, and hence it admits a true probabilistic
interpretation, as opposed to the non-averaged Wigner function.

Notice that in this case $\PB{ H_\mathrm{s}}{F }=0$ and that, for
any phase space function $\Psi=\Psi(\theta,j)$,
\[
\begin{split}
     \oint \frac{\ud\theta}{2\pi} \derp{\Psi}{p} &=
     \oint \frac{\ud\theta}{2\pi} \PB{x}{\Psi}
     =  \oint \frac{\ud\theta}{2\pi} \left( \derp{x}{\theta}\derp{\Psi}{j} -
    \derp{x}{j}\derp{\Psi}{\theta} \right) \\
    &=
     \oint \frac{\ud\theta}{2\pi} \left( \derp{x}{\theta}\derp{\Psi}{j} +
    \derpp{x}{j}{\theta}\Psi \right) =     \dert{}{j} \oint \frac{\ud\theta}{2\pi} \left(\derp{x}{\theta} \Psi
    \right).
\end{split}
\]
Using the last expression we can average \Eqref{FP} to obtain the
following averaged Fokker-Planck equation:
\begin{equation} \label{AvFP}
    \derp{F}{t} = \derp{}{j} \left( \hbar \bar{\mathcal N}
    \derp F j + \bar{\mathcal D} F \right),
\end{equation}
where we have introduced $\bar{\mathcal D}(j)$ and $\bar{\mathcal
N}(j)$, which are defined as follows:
\begin{subequations}
\begin{align}
    \bar{\mathcal D} &=  \oint \frac{\ud \theta}{2\pi}  \derp{x}{\theta}
    \mathcal D
    = -i \sum_{n=-\infty}^\infty |x_n(j)|^2 n D_n(j), \label{mathcalD}\\
    \bar{\mathcal N} &= \oint \frac{\ud\theta}{2\pi} \derp{x}{\theta} \derp {\mathcal N}{\theta}
    =   \sum_{n=-\infty}^\infty |x_n(j)|^2 n^2 N_n(j).
    \label{mathcalN}
\end{align}
\end{subequations}
Eqs.~\eqref{mathcalD} and \eqref{mathcalN} can be further
simplified. Taking into account \Eqref{DnNn} and the fact that
$1/(z+i\epsilon)= \mathrm{P}(1/z) - i \pi \delta(z)$ we can write
\begin{equation} \label{DnSep}
    D_n(j)
    = \mathrm{PV} \int \frac{\ud\omega}{2\pi}
    \frac{2i \tilde D(\omega)}{\omega+n\Omega(j)} + \tilde
    D(n\Omega(j)),
\end{equation}
where we took into account that $\tilde D(\omega)$ is an odd
function. When summing over $n$ in \Eqref{mathcalD}, only the last
term in \Eqref{DnSep} will contribute for every $D_n$. Since the
contributions from the first term in $D_n$ and $D_{-n}$ cancel for
each $n$ because
\[
    \frac{D(\omega)}{\omega - n \Omega(j)} - \frac{D(\omega)}{\omega + n \Omega(j)}
    = \frac{2nD(\omega)}{\omega^2-n^2\Omega^2(j)}
\]
is an odd function and integrates to zero. The final result for
the coefficient $\bar{\mathcal D}$ is
\begin{subequations}
\begin{equation}
    \bar{\mathcal D}(j) = i \sum_{n=-\infty}^{\infty} |x_n(j)|^2 n
    \tilde D (n\Omega).
\end{equation}
Performing similar steps with $N_n$, we get the following
expression for $\bar{\mathcal N}$:
\begin{equation}
    \bar{\mathcal N}(j) = \fud \sum_{n=-\infty}^{\infty} |x_n(j)|^2 n^2
    \tilde N(n\Omega).
\end{equation}
\end{subequations}

Particularizing to the case of an ohmic environment initially at
zero temperature, we have $\tilde N(\omega)=\eta |\omega|$,
$\tilde D(\omega)=i\eta\omega$, which lead to
\begin{subequations}
\begin{align}
    \bar{\mathcal D}(j) &= 2 \eta \Omega(j) \sum_{n=0}^{\infty} |x_n(j)|^2 n^2,     \\
    \bar{\mathcal N}(j) &= \eta \Omega(j) \sum_{n=0}^{\infty} |x_n(j)|^2
    n^3.
\end{align}
\end{subequations}

\section{Environment induced decay rate}

Our aim in this section is to compute the environment induced
decay rate by solving the averaged Fokker-Planck equation,
\Eqref{AvFP}, with the appropriate boundary conditions. We begin
by introducing a simplifying hypothesis.

\subsection{The harmonic approximation}

The non-dissipative dynamics described by \eqref{ClasMot} is
approximately harmonic for energies much lower than the escape
energy $\Es$. In this case the frequency of that motion is simply
$\Omega_0$. For intermediate energies, the motion is qualitatively
similar, but with a somewhat smaller frequency. It is only for
energies extremely close to the escape energy $\Es$ that the
motion is substantially different: the particle needs a very large
amount of time to complete one period and hence the frequency
tends to zero.

In order to obtain an order of magnitude estimate of the escape
rate, it is legitimate to approximate the classical motion of
\Eqref{ClasMot} by its harmonic approximation, since this
approximation is qualitatively valid for all the values of the
energy except for a very small region of energies extremely close
to $\Es$.\footnote{In fact, an exact solution of the equations of
motion, together with a numerical analysis of the Fokker-Plank
equation, reveal that the result that we will obtain here is not
only qualitatively, but also quantitatively valid within the
degree of approximation we are working.}

In the case of vacuum decay in quantum field theory studied by
\cite{CalzettaRouraVerdaguer01,CalzettaRouraVerdaguer02} the
harmonic approximation could not be introduced, since in that case
there existed a frequency threshold with a value greater than
$\Omega_0$ in the dissipation, so that only Fourier modes with a
frequency higher than the threshold contributed to it. Hence it
was crucial to consider the fully non-linear dynamics of the
system. On the other hand, in our case the noise and dissipation
kernels do not exhibit such a threshold and it is possible to
introduce the harmonic approximation, which amounts to neglect the
Fourier modes at higher frequency in front of the lowest ones in
the solution of \Eqref{ClasMot}.

Neglecting the nonlinear term of the potential in
\Eqref{ClasMotb}, the solution of the equations of motion with
final conditions $X(t)=x$ and $P(t)=p$ can be written as
\begin{subequations} \label{HarmMotion}
\begin{align}
    X(t') &= \fud \left( x - \frac{ i p}{M \Omega_0} \right)
    \expp{i\Omega_0(t'-t)} + \fud \left( x + \frac{ i p}{M \Omega_0} \right)
    \expp{-i\Omega_0(t'-t)},\\
    P(t') &= \fud \left( i M \Omega_0 x +  p \right)
    \expp{i\Omega_0(t'-t)} + \fud \left( -i M \Omega_0 x +  p \right)
    \expp{-i\Omega_0(t'-t)}.
\end{align}
\end{subequations}
The frequency of the motion is simply given by $\Omega(j) =
\Omega_0$, and thus the action variable $j$ is simply given by
$j=\varepsilon/\Omega_0$, being $\varepsilon = p^2/(2M) +
\Omega_0^2 x^2 /2$ the energy. In order to determine the action
variable, we compare the solution for $X(t')$ with
\Eqref{Xthetaj}, and write
\begin{equation}
    X(t') = \fud \sqrt{\frac{2j}{M \Omega_0}} \expp{ i [\theta +
    \Omega_0(t'-t)]} + \fud \sqrt{\frac{2j}{M \Omega_0}} \expp{ -i [\theta +
    \Omega_0(t'-t)]}
\end{equation}
where we have identified the angle variable $\theta$ as
\begin{equation}
    \expp{i \theta} = \frac{ M \Omega_0 x - i p}{\sqrt{(M \Omega_0 x)^2
    + p^2}}.
\end{equation}
The decomposition of the $x$ variable in terms of $\theta$ and $j$
is given by
\begin{equation}
    x(\theta,j)=\frac{1}{2} \sqrt{\frac{2j}{M\Omega_0}} \expp{i\theta}
    + \frac{1}{2} \sqrt{\frac{2j}{M\Omega_0}} \expp{-i\theta},
\end{equation}
so that
\begin{equation}
    x_1(j)=x_{-1}(j)=\frac{1}{2} \sqrt{\frac{2j}{M\Omega_0}},
    \qquad x_n = 0, \quad n \neq -1,1.
\end{equation}
The coefficients $\bar{\mathcal D}(j)$ and $\bar{\mathcal N}(j)$
can be therefore expressed as
\begin{equation}
    \bar{\mathcal D}(j) =  2 \gamma  j, \qquad \bar{\mathcal N}(j) =
    \gamma j,
\end{equation}
[we recall that $\gamma \equiv \eta/(2M)$], and the averaged
Fokker-Planck equation may be written as
\begin{equation}
    \derp{F}{t} = 2\gamma \derp{}{j} \left( \frac {\hbar
    \Omega_0}{2} j \derp{F}{j} + \Omega_0 j F \right),
\end{equation}
or, equivalently, working with energies,
\begin{equation}\label{FPHarmonic}
    \derp{F}{t} = 2 \gamma \derp{}{\varepsilon} \left( \Eo \varepsilon \derp{F}{\varepsilon} + \varepsilon F \right),
\end{equation}
where $\Eo \equiv \hbar \Omega_0 /2$. It can also be rewritten as
a conservation equation,
\begin{gather}
    \derp{F}{t} + \derp{\Phi}{\varepsilon} = 0, \label{Conservation}\\
\intertext{where}
    \Phi = - 2 \gamma \left( \Eo \varepsilon \derp{F}{\varepsilon} + \varepsilon F
\right)\label{Phi}
\end{gather}
is the probability flux.

\subsection{Escape rate: Normal mode analysis}

Assuming that the Fokker-Planck equation can be decomposed into a
sum of normal modes,
\begin{equation}
    F(t,\varepsilon) = \sum_r c_r \expp{-r t} f_r(\varepsilon),
\end{equation}
we get the following time-independent equation:
\begin{equation} \label{FPTI}
    L f_r + r f_r = 0, \qquad
    L = 2 \gamma \dert{}{\varepsilon} \left( \Eo \varepsilon \dert{}{\varepsilon} + \varepsilon
    \right).
\end{equation}
The boundary conditions of the partial differential equation are
the following. First, we have assumed that the particle is removed
once it arrives at the separatrix, so that there will be no
probability to find the particle at the separatrix: $f_r(\Es)=0$.
Secondly, we will also assume a vanishing flux of incoming
particles at $\varepsilon=0$, \ie, $\Phi(0)=0$, which is
equivalent to demanding the finiteness of $f_r$ and its derivative
at $\varepsilon=0$ ($f_r(0),f'_r(0)<\infty$), as can be seen from
\Eqref{Phi} .

The normal-mode analysis can be formulated as a standard
Sturm-Liouville problem. The operator $L$, which may be written as
\begin{equation}
    L = 2 \gamma \expp{ - \varepsilon/\Eo} \Eo \dert{}{\varepsilon} \left( \expp{
    \varepsilon/\Eo }\varepsilon\dert{}{\varepsilon}
    \right) + 2 \gamma,
\end{equation}
is self-adjoint with the aforementioned boundary conditions and
the scalar product defined by
\begin{equation}
    (f,g) = (2 \gamma)^{-1} \int_0^{\Es} \ud \varepsilon \expp{\varepsilon/\Eo} f^*(\varepsilon) g(\varepsilon).
\end{equation}
Thus, the theory of differential equations guarantees that the
eigenfunctions $f_r(\varepsilon)$ constitute a complete orthogonal
set, and that the eigenvalues $r$ are real \citep{CourantHilbert}.
Furthermore, the operator $L$ is negative definite, which can be
seen as follows:
\[
\begin{split}
    (f,Lf) &=\Eo \int_0^{\Es} \ud \varepsilon \expp{\varepsilon/\Eo} f^*(\varepsilon) \dert{}{\varepsilon}
    \left[ \varepsilon \expp{-\varepsilon/\Eo} \dert{}{\varepsilon} \left( \expp{\varepsilon/\Eo}f(\varepsilon)
    \right) \right] \\
    &= - \Eo \int_0^{\Es} \ud \varepsilon \varepsilon \expp{-\varepsilon/\Eo} \left| \dert{}{\varepsilon}
    \left(\expp{\varepsilon/\Eo} f(\varepsilon) \right) \right|^2 < 0, \quad  f \neq
    0,
\end{split}
\]
where we have integrated by parts in the last equality. This
implies that the eigenvalues $r$ are always positive, as expected.

Performing the change of variables $f_r(\varepsilon) = 2 \gamma
\expp{-y} \bar f_{\bar r}(y)$, where $y\equiv \varepsilon/\Eo$ and
$\bar r \equiv r/(2\gamma)$, the differential equation $L f_r = r
f_r$ adopts the form of the Laguerre differential equation,
\begin{equation} \label{LaguerreDiff}
    y \bar f_{\bar r}''(y) + (1-y) \bar f'_{\bar r}(y) + \bar r f_{\bar r}(y) =
    0,
\end{equation}
whose unique regular solution is given by
\begin{equation}
    \bar f_{\bar r}(y) = N L_{\bar{r}}(y),
\end{equation}
where $L_{\bar{r}}(y)$ are Laguerre functions (which reduce to the
Laguerre polynomials in the case of non-negative integer $\bar r$;
see \citealp{GradsteynRyzhik}), and $N$ is a normalization
constant.

The solution we have found verifies the first of the boundary
conditions, the regularity at the origin. Now we impose the second
of the boundary conditions, namely $\bar f_{\bar r}(y_\mathrm s) =
0$ (with $y_\mathrm s \equiv \Es/\Eo$). This boundary condition
will imply a discretization on the possible values of the escape
rate $r$. Finally, knowing the initial state
$\fini(\varepsilon)=F(0,\varepsilon)$, we will be able to
reconstruct the solution by computing the coefficients $c_r$:
\begin{equation}
    c_r = (f_r,\fini) = N \int_0^\Es \ud \varepsilon L_{\frac{r}{2\gamma}}\left( \frac{\varepsilon}{
    \Eo} \right) \fini(\varepsilon).
\end{equation}

Unfortunately, it is not possible to determine analytically the
possible values of $\bar r$ from the equation
$L_{\bar{r}}(y_\mathrm s)=0$. However, if the potential barrier
were infinitely high, the second boundary condition would read $
    \lim_{y\to\infty}  \expp{-y} L_{\bar r}(y) = 0
$, implying that $L_{\bar r}(y)$ should have, at most, a
polynomial behavior at infinity, and the eigenvalues would be
$\bar r=0,1,2\ldots$, so that eigenmodes with $\bar r \neq 0$
would decay in a time given by $\gamma^{-1}$ or shorter. However,
the potential barrier corresponds to a some large but finite
energy, and hence the real eigenvalues differ from those computed
in the infinite barrier case by a small quantity, at least for
those eigenvalues corresponding to eigenstates with characteristic
energies much lower than the potential barrier. Thus, we can
compute them perturbatively.

Furthermore, the relevant contribution to the decay rate will be
given by the mode with the lowest eigenvalue, which will fulfill
the condition $\bar r \ll 1$, since the remaining modes will decay
in a time of order $\gamma^{-1}$ at most. Thus, we proceed to
 compute perturbatively the lowest order mode by expanding the
Laguerre function around $\bar r=0$:
\newcommand{\gammaE}{\gamma_{\scriptscriptstyle \mathrm{E}}}
\begin{equation} \label{LaguerreL}
    L_{\bar{r}}(y) = 1 + \bar r [ \ln y + \gammaE -
    \Ei(y)] + O(\bar r^2),
\end{equation}
where $\gammaE=0.577216\ldots$ is Euler's constant, and $\Ei(y)$
is the exponential integral defined as $\Ei(y)\equiv \mathrm{PV}
\int_{-\infty}^y ({\expp {u}}/{u}) \vd u$. The expansion in
\Eqref{LaguerreL} can be found by solving perturbatively
Eq.~\eqref{LaguerreDiff} up to order $\bar r$, and imposing the
correct boundary condition at $y=0$.

Therefore, imposing the boundary condition $L_{\bar r}(y_\mathrm
s) = 0$ is equivalent to demanding
\begin{equation}
    r = 2 \gamma \bar r \approx   \frac{2 \gamma}{\Ei(\Es/\Eo)-\ln (\Es/\Eo) - \gammaE}.
\end{equation}
Since $\Es$ is much larger than $\Eo$, the exponential integral
can be approximated by $\Ei(\Es/\Eo) \approx
(\Eo/\Es)\expp{\Es/\Eo}$, and the other two terms in the
denominator become negligible in front of this one. Hence, we may
approximate the lowest order solution by
\begin{equation} \label{EscapeRate}
    r \approx \frac{2 \gamma \Es}{\Eo} \exp{ \left( -
    \frac{\Es}{\Eo}\right)}.
\end{equation}
Equation~\eqref{EscapeRate}, which gives the probability per unit
time for a particle to jump the barrier, is our final result for
the escape rate.

In the case of vacuum decay in quantum field theory studied by
\cite{CalzettaRouraVerdaguer01,CalzettaRouraVerdaguer02} the
time-independent Fokker-Planck equation had a continuous spectrum,
and therefore in that case it was not possible to perform either
an eigenvalue expansion, like the one we have performed in this
subsection, or follow Kramers's method, as will be done in the
next subsection.

\subsection{Escape rate: Kramers's method}

Our Fokker-Planck equation \eqref{FPHarmonic} is completely
analogous to that found by \cite{Kramers40} in the classical
underdamped case, once we replace $k T$, the temperature times
Boltzmann's constant, by the zero point energy of the harmonic
oscillator $\Eo = \hbar \Omega_0/2$.\footnote{Notice that this is
only true under the harmonic approximation. Had not we neglected
the cubic term in the equations of motion, our Fokker-Planck
equation would not be equivalent to the one found by Kramers in
the classical underdamped case.} Therefore, as an alternative to
the previous subsection, we can apply the same method as Kramers
in order to compute the decay rate.

Instead of imposing the correct boundary conditions, we look for
the solutions of the Fokker-Planck equation with constant flux
$\Phi=\Phi_0$. Since the flux can be rewritten as
\begin{equation}
    \Phi =
    - 2 \gamma \expp{-\varepsilon/\Eo} \varepsilon \Eo \derp{}{\varepsilon}\left(\expp{\varepsilon/\Eo} F \right),
\end{equation}
these solutions will be given by
\begin{equation}
\begin{split}
    F(\varepsilon) &=  \frac{\Phi_0}{2 \gamma \Eo} \expp{-\varepsilon/\Eo} \int_{\varepsilon}^{\Es} \ud {\varepsilon'}
    \frac{\expp{\varepsilon'/\Eo}}{\varepsilon'} \\
    &= \frac{\Phi_0}{2 \gamma \Eo}  \expp{-\varepsilon/\Eo} \left[
    \Ei(\Es/\Eo) - \Ei(\varepsilon/\Eo) \right].
\end{split}
\end{equation}
Again, we have imposed that $F(\Es)=0$ because we assume that when
a particle arrives at the separatrix it never reenters the
potential well region. Notice that the solution we have found has
a logarithmic singular behavior at small energies, which takes
into account the injection of a probability flux through the point
$\varepsilon=0$ necessary for the maintenance of the constant
flux. However, since we expect the flux of probability to be very
small, this contribution will be not very significative and will
affect only the region of energies $\varepsilon \lesssim \Eo$,
which are much lower than the scape energy $\Es$.

We can compute the flux $\Phi_0$ by imposing the correct
normalization of the averaged Wigner function $F(\varepsilon)$,
namely $\int_0^{\Es} \ud \varepsilon F(\varepsilon) = 1$:
\begin{equation}\label{NormPhi}
    \frac{\Phi_0}{2 \gamma \Eo} \int_0^{\Es} \ud \varepsilon \expp{-\varepsilon/\Eo} \int_{\varepsilon}^{\Es} \ud
    {\varepsilon'}
    \frac{\expp{\varepsilon'/\Eo}}{\varepsilon'} = 1.
\end{equation}
The main contribution to the integral over $\varepsilon'$ is due
to those values for $\varepsilon'$ which differ from $\Es$ by a
quantity of order $\Eo$. We may replace $\varepsilon'$ by its
value at the separatrix, $\Es$. Making this approximation we can
perform analytically the integrals in \Eqref{NormPhi}. Retaining
only those terms which have an exponential factor
$\expp{\Es/\Eo}$, we obtain the final result for the flux of
particles:
\begin{equation}
    \Phi = \Phi_0 \approx \frac{2 \gamma \Es}{\Eo} \exp \left( - \frac{\Es}{\Eo}
    \right).
\end{equation}
Although we have assumed that the flux is constant in order to
solve the differential equation, this is not actually the case,
and the flux at the separatrix $\Phi_\mathrm s$ is proportional to
the total probability of finding the particle in the potential
well region: $\Phi_\mathrm s(t) = P(t) \Phi_0$. Integrating the
conservation equation \eqref{Conservation} with respect to the
energy, we easily find that the total probability decay follows
the law
\begin{equation*}
    \dert{P(t)}{t} + P(t) \Phi_0 = 0,
\end{equation*}
which may be integrated to give $P(t) = \expp{ - \Phi_0 t}$. Hence
the probability flux $\Phi_0$ is to be identified with the escape
rate $r$ of last subsection. We see that both methods agree.

\section{Discussion}

The study of quantum tunneling based on a real time formulation
seems crucial in order to address highly non-equilibrium
situations in which no adiabaticity assumptions can be made. As
mentioned in the introduction, a first step in this direction was
made in \cite{CalzettaRouraVerdaguer01,CalzettaRouraVerdaguer02},
where the effect on vacuum decay in quantum field theory due to
the backreaction of the short-wavelength modes was analyzed by
regarding the long-wavelength modes responsible for tunneling as
an open quantum system. In that case it was found that such a
backreaction seemed to yield an enhancement of the decay rate.
There are, however, a couple of aspects which deserve, in our
opinion, a more careful analysis. The first one is the need for a
suitable identification of the tunneling degrees of freedom and
the corresponding implementation of a system-environment
separation which leads to the reformulation of the problem in
terms of an open quantum system. Secondly, when solving the master
equation that governs the time evolution of the reduced Wigner
function for the tunneling degree of freedom in order to obtain
the vacuum decay rate, the attention was focused on the
backreaction of the environment (the short-wavelength modes) and
the higher derivative terms which would be uniquely responsible
for tunneling if the system were isolated were neglected.

In this paper we have considered a fairly simple quantum
mechanical open system, in which the system-environment separation
is given before-hand and the coupling constant governing the
interaction between the system and the environment can be adjusted
at will, rather than being self-consistently determined, as
happened to be the case for vacuum decay in field theory.
Therefore, it is interesting to check whether a similar
enhancement of the tunneling rate is obtained in that simpler
model and see if such an effect is generic. Furthermore, the
result can be considered more robust than that in the field theory
case since the number of important assumptions made is rather
small. It is, thus, worthwhile to elaborate on this point and
recall the different approximations employed throughout the paper
in order to obtain the decay rate.

First of all, the master equation that we are considering in this
paper can be obtained by keeping terms of order $\hbar^2$ and
$\hbar\gamma$ at most, and restricting to values of $\gamma$ which
are small enough (in particular we should take $(\gamma/\Omega_0)
S \lesssim \hbar$, where $S$ is the characteristic action of the
problem), so that the terms of order $\gamma^2$ or higher can be
neglected\footnote{One might be concerned that the truncation of
higher orders in $\gamma$ when considering times much larger than
the characteristic relaxation timescale (\ie, $\gamma t \gg 1$),
as required in order to compute the decay rate, might no longer be
valid due to the existence of secular terms among the terms of
higher order in $\gamma$ that have been neglected. Although
arguments that justify such a truncation when computing the decay
rate can be given, this point might deserve a more careful
analysis.} \citep{RouraThesis}. On the other hand, the master
equation obtained in this way contains a term with third order
derivatives of the reduced Wigner function with respect to the
momentum (the term of order $\hbar^2$). This sort of terms, which
are absent in any diffusion equation with a finite Kramers-Moyal
expansion associated to a classical stochastic process, are
intimately related to genuinely quantum effects due to the
nonlinearities of the potential and imply that even a reduced
Wigner function which is initially positive everywhere will
acquire negative values when it evolves in time. Moreover, this
higher derivative term would be uniquely responsible for tunneling
if the system where isolated. Despite its remarkable features,
this term has been neglected in the present article, since
otherwise we were unable to obtain analytical results for the
decay rate. This approximation, whose justification will be
further discussed below, constitutes the most drastic assumption
made throughout the paper.

Having neglected the higher derivative terms, the master equation
is equivalent to the Fokker-Planck equation associated to a
Langevin equation with a Gaussian stochastic source characterized
by a nonlocal correlation function (the nonlocal noise kernel).
From this point on, most of the approximations employed in order
to compute the decay rate are more or less standard
\citep{HanggiEtAl90}. First, we change to action-angle variables
and make use of an adiabatic approximation to eliminate the fast
variable (the angle). This is consistent provided that $\gamma \ll
\Omega_0$, which is in agreement with the previous assumption of
small enough values for $\gamma$. Next, a harmonic approximation
is introduced for the solutions to the equations of motion for the
isolated system which appear in both the master equation and the
Fokker-Planck equation. This approximation helps to obtain a
rather simple result for the decay rate and can be justified both
qualitatively and quantitatively, in contrast to the field theory
case analyzed by
\cite{CalzettaRouraVerdaguer01,CalzettaRouraVerdaguer02}, where
the existence of a threshold for the dissipation and noise kernels
would preclude such an approximation. Finally, it is assumed that
the characteristic decay time is much smaller than the relaxation
and dynamical timescales: $r \ll \gamma \ll \Omega_0$, where $r$
is the decay rate (the inverse of the decay time). Although
$\gamma$ was required to be small enough, the previous inequality
can be fulfilled provided that the potential barrier is
sufficiently high, \ie, $\Es$ is large enough.

After reviewing the main approximations employed, let us compare
the result obtained for the decay rate to that of
\cite{CaldeiraLeggett81,CaldeiraLeggett83a} as well as to the
tunneling rate when the coupling to the environment is absent.
Whereas Caldeira and Leggett found that the interaction with the
environment tends to suppress tunneling, we are not going to
recover that result since, having neglected the higher derivative
terms responsible for tunneling in an isolated system, the
dissipation and diffusion terms that appear in the master equation
due to the backreaction of the environment will always lead to a
positive (or, at most, vanishing) probability of escaping from the
metastable well. Nevertheless, if the decay rate obtained were
much smaller or much larger than the tunneling rate for the
isolated system so that the timescales governing both processes
are very different, one could expect that the contribution to  the
decay rate from the process with a shortest characteristic
timescale would be dominant.

The tunneling rate for an isolated system initially trapped in the
metastable minimum of the potential considered in this paper is
\citep{CaldeiraLeggett83a}
\begin{equation}
r_\text{t} \sim  \Omega_0 \left(\frac{\Es}{\Eo}\right)^{1/2}
\exp\left(-\frac{18}{5} \frac{\Es}{\Eo}\right) \label{Tunel},
\end{equation}
where we recall that $\Eo$ is the zero-point energy of a harmonic
oscillator of frequency $\Omega_0$. On the other hand,
\cite{CaldeiraLeggett81,CaldeiraLeggett83b} obtained the
modification of the tunneling rate due to the interaction with the
environment, which in the case of small dissipation is given by
\begin{equation}
r_\mathrm{t} \sim  \Omega_0 \left(\frac{\Es}{\Eo}\right)^{1/2}
\exp\left[- \frac{\Es}{\Eo} \left( \frac{18}{5} + \frac{54
\zeta(3)}{\pi^3} \frac{\gamma}{\Omega_0} \right)\right].
\label{TunelEnv}
\end{equation}
Therefore, since the interaction with the environment simply adds
a negative contribution to the exponent, it always tends to
suppress the tunneling rate. Finally, the decay rate due to the
activation-like effect obtained in the previous section
corresponds to
 \begin{equation}
r \sim \gamma \frac{\Es}{\Eo} \exp{\left(-\frac{\Es}{\Eo}\right)}
\label{Activation},
\end{equation}
which is valid for small $\gamma$ and large $\Es$. When $\gamma$
is very small the decay rates  from Eqs.~(\ref{Tunel}) and
(\ref{TunelEnv}) become very close. Furthermore, although a small
$\gamma$ implies a small contribution to the decay rate associated
to the activation-like process, it can always be made arbitrarily
larger than the tunneling rate by taking $\Es$ large enough.

The fact that the activation-like decay rate can be made
arbitrarily large as compared to the tunneling rate for the
isolated system seems to suggest, as mentioned above, that even if
the two phenomena were considered simultaneously, the former would
be expected to dominate. If that were the case, it would imply
that the usual instanton methods, when applied to a system
interacting with an environment, downplay the role of the
backreaction of the environment on the system dynamics. This can
be qualitatively interpreted in the following way: while, roughly
speaking, the tunneling effect for an isolated system can be
regarded as a consequence of the energy fluctuations implied by
Heisenberg's uncertainty principle, the interaction with an
environment would induce fluctuations on the system due to the
quantum fluctuations of the environment itself
\citep{NagaevButtiker02}, which would enhance the tunneling rate.

Nevertheless, in order to reach a definite conclusion it would be
desirable to deal with the two contributions simultaneously and
make sure that the higher derivative terms do not disrupt the
effect of the backreaction terms, even when the timescale for the
contribution to the decay rate from the former terms is much
longer than the timescale associated to the activation-like
process. Unfortunately, dealing with the higher derivative terms
is not an easy task, and it seems hard to provide a real time
description of the tunneling process in terms of the Wigner
function even for an isolated system; see, however,
\cite{RiskenVogel88} for a first step in that direction. One may
try to gain partial information on the relation between the two
processes by considering different potentials with wider or
narrower barriers, since one would naively expect that tunneling
is suppressed for wide barriers while the activation-like
contribution does not depend so much on the width of the barrier,
as long as the height remains the same. A couple of comments
concerning the freedom to modify the system potential are,
nevertheless, in order. First, the potential must be analytic in
order the derive the master equation for the reduced Wigner
function. Second, when solving the formally equivalent problem of
a classical Brownian particle escaping from the potential well, we
assumed that once the particle reaches the maximum of the barrier
it escapes and never comes back, but if a very wide barrier is
considered, the probability that the particle comes back due to
the fluctuations may become no longer negligible.

We close this section insisting on the importance of finding a
satisfactory method to deal with the higher derivative terms,
which would be very helpful in order to elucidate whether the
enhancement of the decay rate obtained in this paper and entirely
due to the back reaction from the environment fluctuations would
still persist when the terms responsible for tunneling in isolated
systems are also taken into account. Such a method would have an
interest in its own right even if the results of
\cite{CaldeiraLeggett81,CaldeiraLeggett83a} were finally recovered
when properly taking into account the higher derivative terms,
since it would constitute a key step in formulating a real time
description of tunneling.

\section*{Acknowledgments}

\hyphenation{CONICET} We would like to thank Bei-Lok Hu, Fernando
Lombardo, Slava Mukhanov and Renaud Parentani for interesting
discussions in Peyresq. D.~A. acknowledges support of a FI grant
from the Generalitat de Catalunya, and A.~R. acknowledges support
from NSF under Grant~PHY-9800967. This work is also supported in
part by MICYT Research Project No.~FPA-2001-3598, European Project
HPRN-CT-2000-00131, ANPCYT through Project No.~PICT99 03-05229,
CONICET, UBA and Fundacion Antorchas.


\end{document}